\documentclass[twocolumn,a4paper,10pt]{article}
\usepackage{amsmath,amsfonts,amssymb}
\usepackage{epsfig}
\usepackage{times}
\usepackage{epsfig}
\usepackage[small,hang]{caption}
\usepackage{natbib}
\usepackage{graphicx}
\usepackage{xcolor}
\usepackage{bm}
\usepackage{accents}
\usepackage{subfigure}
\usepackage{afterpage}

\newlength{\dhatheight}

\makeatletter

\newcounter{numbersec}
\renewcommand{\section}[1]{\par\noindent\stepcounter{numbersec}
\par
\vspace{6pt}
\noindent\textbf{\large   \arabic{numbersec} \hspace*{0.3cm} #1 }
\par
\vspace{2pt}
}
\renewcommand{\subsection}[1]{
\par
\vspace{6pt}
\noindent\textbf{#1}
\par
}
\renewcommand{\subsubsection}[1]{%
\par
\vspace{6pt}
\textbf{#1.}
}

\newcommand{\Abstract}{\par\vspace{6pt}\noindent\textbf{\large Abstract}\par\vspace{2pt}}
\newcommand{\Acknowledgments}{\par\vspace{6pt}\noindent\textbf{\large Acknowledgments }\par\vspace{2pt}}
\newcommand{\References}{\par\vspace{6pt}\noindent\textbf{\large References }\par\vspace{2pt}}

\makeatother

\title{\vspace*{-12mm}
\LARGE \sc \textbf{  
Effects of sweeps and ejections on amplitude   \\
modulation in a turbulent channel flow
}}
\author{ \Large \bf \textit{ 
A. Andreolli$^{1}$, D. Gatti$^{1}$, R. Vinuesa$^{2}$, R. \"Orl\"u$^{2}$ and P. Schlatter$^{2}$ }  \\ \\
\bf  $^{1}$ \textit{ Institute of Fluid Dynamics, Karlsruhe Institute of Technology, Kaiserstra\ss e 10} \\
\bf \textit{76131 Karlsruhe, Germany}\\
\bf  $^{2}$ \textit{ SimEx/FLOW, Engineering Mechanics, KTH Royal Institute of Technology} \\
\bf  \textit{SE-100 44 Stockholm, Sweden} \\ \\
\underline{\bf a.andreolli@kit.edu}
}
\date{}

\begin{document}

\maketitle
\thispagestyle{empty}

\Abstract

Conditional averages are used to evaluate the effect of sweeps and ejections on amplitude modulation. This is done numerically with a direct numerical simulation (DNS) of a channel flow at friction Reynolds number $Re_\tau = 1000$ in a minimal steram-wise unit (MSU). The amplitude-modulation map of such DNS is also compared to the one of a regular channel flow in a longer streamwise domain (LSD), in order to assess its validity for this study. The cheaper MSU is found to provide a good representation of the modulation phenomena in the LSD.
As for conditional averages, the amplitude-modulation coefficient is conditioned on the sign of the large-scale fluctuations. Care must be exerted in defining such a coefficient, as the conditioned large-scale fluctuation has non-zero average, indeed as a consequence of conditioning. Both sweeps and ejections (positive and negative large-scale fluctuation events) are found to have a positive contribution to amplitude modulation in the buffer layer, and a negative one in the outer layer. The negative-modulation region is found to shrink in case of ejections, so that the positive-modulation region extends farther away from the wall.
Two more conditional statistics are used to provide an alternative representation of amplitude modulation and insights into the characteristics of the large-scale structures.

\section{Introduction}

Velocity statistics in wall turbulence are well known to scale in viscous units \citep{pope-2000}; viscous scaling is meant as a non-dimensionalisation with either the friction velocity $u_\tau = \sqrt{\tau_w/\rho}$ ($\tau_w$ being the mean wall shear stress, $\rho$ the density) or its corresponding length scale $\ell = \nu/u_\tau$. All quantities in this paper are scaled in such way. This scaling however fails e.g. for all Reynolds normal stresses due to the arisal of (very-)large-scale motions at high Reynolds number \citep{hutchins-marusic-2007-b}, which interact with the smaller, near-wall ones, distorting them. Such interaction phenomena are usually depicted in terms of large-scale superposition, amplitude modulation and frequency modulation \citep{baars-hutchins-marusic-2016}.

Large scales usually have a high energetic imprint in the outer section of the flow, away from the wall \citep{lee-moser-2015}; nevertheless, their presence also excites the low wave-number modes in the wall proximity \citep{hutchins-marusic-2007}. This phenomenon goes under the name of superposition; superposition is also thought to be responsible for the two remaining phenomena, namely amplitude and frequency modulation \citep{baars-hutchins-marusic-2016,agostini-leschziner-2019}. The velocity signal of a flow can be decomposed into a small- and a large-scale part by appropriate filtering; by doing so, it is found that the amplitude of the small-scale signal appears to increase when the large-scale fluctuation is positive, whereas it decreases when large scales experience a negative fluctuation \citep{mathis-hutchins-marusic-2009}. The same applies to frequency \citep{baars-hutchins-marusic-2016}.

As in \cite{dogan-etal-2019}, many filtering techniques are available for the scale decomposition of the signal, either based on Fourier modes or data-driven approaches such as the Empirical Mode Decomposition (EMD); moreover, many quantities based on either correlations or variances have been defined in literature to reveal the presence of amplitude modulation. The same authors report no qualitative difference among the results yielded by each of these approaches.

Traditionally, amplitude modulation has been investigated by the means of single-point correlations \citep{mathis-hutchins-marusic-2009}; by doing so, amplitude modulation is found to occur in the near-wall region up until the buffer layer ($y \approx 10$, $y$ being the wall-normal coordinate in viscous units), where the amplitude-modulation correlation is positive; the correlation then changes sign in the logarithmic layer to show negative values in the outer layer. Thus in the outer layer small scales are reduced in amplitude in presence of positive large-scale fluctuations and vice versa, reverting the near-wall scenario; this has been linked to both intermittency \citep{mathis-hutchins-marusic-2009} and the sign of the large-scale velocity gradient, which then interferes with the production of energy at small scales \citep{agostini-leschziner-2019}. One-point amplitude-modulation coefficients are intimately related to the skewness of the velocity probability distribution function \citep[PDF, see][]{mathis-etal-2011}, to the point that the physical interpretation of such one-point statistics as measures of amplitude modulation has been questioned \citep{schlatter-orlu-2010b}. This issue can be worked around by using two-point, non-local correlations (or covariances) to quantify the phenomenon, as in \cite{bernardini-pirozzoli-2011}. By computing the covariance between the large-scale signal and the envelope of the small scales at different wall-normal positions, the authors detect a second, off-diagonal positive peak in the amplitude-modulation covariance map, in addition to the commonly observed diagonal one corresponding to the one in one-point statistics. Such a second peak indicates that the envelope of small scales in the buffer layer ($y \approx 10$) correlates well with the large-scale signal farther away from the wall, namely at $y\approx 100$. Moreover, the second peak appears not to be related to the skewness of the velocity signal, so that its physical interpretation is (hopefully) unambiguous. \cite{agostini-etal-2016} have questioned this two-peaks interpretation, still reporting a wide, relatively flat positive correlation region extending away from the diagonal, in substantial agreement with \cite{bernardini-pirozzoli-2011}.

One more intrinsical feature of amplitude modulation is its asymmetry with respect to sweeps or ejections \citep{agostini-leschziner-2014}. Ejections of low-speed fluid from the near-wall region towards the channel core are associated with negative large-scale fluctuations, whereas sweeps of high-momentum fluid from the core to the wall involve positive large-scale fluctuations; one-point variances and the PDF of small-scale velocity change asymmetrically with respect to positive or negative large-scale fluctuations. 

In this paper, we investigate the effects of such an asymmetry directly on the amplitude-modulation map. To do this, we define a two-point amplitude-modulation covariance analogous to the one of \cite{bernardini-pirozzoli-2011}; then, we calculate its value conditioned on the sign of the large-scale fluctuation. Additional understanding is gained by conditionally computing further statistics, namely two-points covariances of the large- and small-scale signals.

\section{Methodology}

This investigation is performed on a direct numerical simulation (DNS) of an incompressible turbulent channel flow at a friction Reynolds number $Re_\tau=1000$, where $Re_\tau = h u_\tau / \nu$ and $h$ is the channel half-height. Conditional amplitude-modulation maps have been calculated on a minimal streamwise unit (MSU) simulation \citep{abe-antonia-toh-2018} with a streamwise domain length $L_x = 0.4 h$ and a spanwise width of $L_z = 2 \pi h$. MSUs have several advantages with respect to conventional simulations (here referred to LSD, as in long streamwise domain). First and foremost, they are consistently less expensive (both in terms of computation and postprocessing); moreover, outer-layer large-scale structures are energised with respect to the LSD case, so that large-small scale interactions are more pronounced and easier to detect. Finally, large-scale structures become essentially two-dimensional, appearing in the outer layer in anti-symmetric pairs. This orderly pattern contrasts with the meandering of LSDs, and simplifies the discussion of the role of such large scales. Comparing the two cases allows for a discussion of the effects of meandering.

As for the quantification of amplitude modulation, we use a  scale-decomposed two-point skewness $C_{AM}^\ast$ \citep{eitel-amor-etal-2014,mathis-etal-2011,bernardini-pirozzoli-2011}:
\begin{equation}
C_{AM}^\ast = \langle u_{SS}^2(y_{SS}) u_{LS}(y_{LS}) \rangle \, ,
\label{eq:cam}
\end{equation}
where $u_{SS}(y_{SS})$ is the high-pass-filtered streamwise velocity fluctuation signal at wall-normal position $y_{SS}$, whereas $u_{LS}(y_{SS})$ the low-pass-filtered equivalent at a second wall-normal position $y_{LS}$. Both signals are evaluated at the same streamwise and spanwise positions, i.e. no streamwise shifting is used. The filtering is carried out in the spanwise statistically homogeneous direction; based on previous investigations \citep{eitel-amor-etal-2014,dogan-etal-2019} and on the scrutiny of current data, a sharp Fourier filter with a cutoff wavelength of $\lambda_{z,c} = 500$ is used.

Such amplitude-modulation covariance is then conditioned on the sign of the large-scale signal, $u_{LS}({y_{LS}})$, meaning that it is calculated on a sample in which $u_{LS}({y_{LS}})$ has always the same sign. This leads to the definition of two conditioned $C_{AM}^+$ and $C_{AM}^-$, corresponding respectively to sweeping events ($u_{LS}(y_{LS}) > 0$) and ejections ($u_{LS}(y_{LS})<0$). Practically speaking, the conditioning effect is obtained by setting to zero elements of the sample that do not fulfill the condition; by letting $f$ be a generic function, its conditional average is defined as:
\begin{align}
  \langle f(y_{SS}, y_{LS}) \,|\, u_{LS}(y_{LS}) > 0 \rangle = \notag\\
  \langle f(y_{SS}, y_{LS}) \, k_{p}(y_{LS}) \rangle \, , \notag\\[10pt]
  k_{p}(y_{LS}) = \begin{cases}
    1 \;\;\;\;\;\;\text{if }\;u_{LS}(y_{LS}) > 0\text{;} \\
    0 \;\;\;\;\;\;\text{otherwise.}
  \end{cases}
  \label{eq:def_cond_avg}
\end{align}
Averages on negative events are similarly defined, naturally with suited adjustments on signs of the inequality.

From now on, the dependence of fluctuation signals on the spatial coordinate will be dropped for the sake of readability; unless explicitly specified, the signals $u_{SS}$ and $u_{LS}$ are evalued at $y_{SS}$ and $y_{LS}$ respectively, as in equation \eqref{eq:cam}.
Care must be exerted in defining the conditioned $C_{AM}^+$ and $C_{AM}^-$ covariances: suppose, for instance, that one conditioned expression \eqref{eq:cam} as it is. All of the $u_{LS}$ values in the sample would have the same sign, implying that $\langle u_{LS} | u_{LS} > 0 \rangle$ would be positive (and $\langle u_{LS} | u_{LS} < 0 \rangle$ negative), as opposed to the fact that the unconditioned $u_{LS}$ fluctuation signal has zero average by definition. Since $u_{SS}^2$ has a positive average regardless of the sign of conditioning, $C_{AM}^+$ would have an unphysical positive bias given by $\langle u_{SS}^2 \,|\, u_{LS} > 0 \rangle\langle u_{LS} \,|\, u_{LS} > 0\rangle$; instead, $C_{AM}^-$ would have a negative one. This can be circumvented by adopting the following definitions:
\begin{align}
  C_{AM}^+ = \left\langle \left(u_{SS}^2 - \langle u_{SS}^2\rangle\right) u_{LS} \,|\, u_{LS} > 0 \right\rangle \, , \label{eq:camplus}\\
  C_{AM}^- = \left\langle  \left(u_{SS}^2 - \langle u_{SS}^2\rangle\right) u_{LS} \,|\, u_{LS} < 0 \right\rangle \, . \label{eq:camminus}
\end{align}
Notice that both of the factors appearing in \eqref{eq:camplus} and \eqref{eq:camminus} still have non-zero conditional average; nevertheless, the sign of the product of their averages is now physically representative of modulation. Moreover, the two definitions satisfy
\begin{equation}
  C_{AM}^\ast = C_{AM}^+ + C_{AM}^-
\end{equation}
thanks to the definition \eqref{eq:def_cond_avg} of the conditional average, so that $C_{AM}^+$ can be interpreted as the effect of sweeping events on the total amplitude-modulation covariance (and $C_{AM}^-$ the effect of ejections).

Further insights can be gained by considering the conditional autocovariances of $u_{LS}$ and $u_{SS}^2$, namely
\begin{equation}
  \left\langle u_{LS}(y_{SS}) u_{LS}(y_{LS}) \,|\, u_{LS}(y_{LS}) \gtrless 0 \right\rangle
\end{equation}
and
\begin{equation}
  \left\langle u_{SS}^2(y_{SS}) u_{SS}^2(y_{LS}) \,|\, u_{LS}(y_{LS}) \gtrless 0 \right\rangle;
\end{equation}
note the wall-normal position at which the signals are evaluated. The former provides information about the structure of large-scale fluctuations. The latter is a different way of investigating amplitude modulation: positive large-scale fluctuations are expected to correspond to an increased small-scale activity, and vice versa.

\begin{figure}
  \centering
  \includegraphics{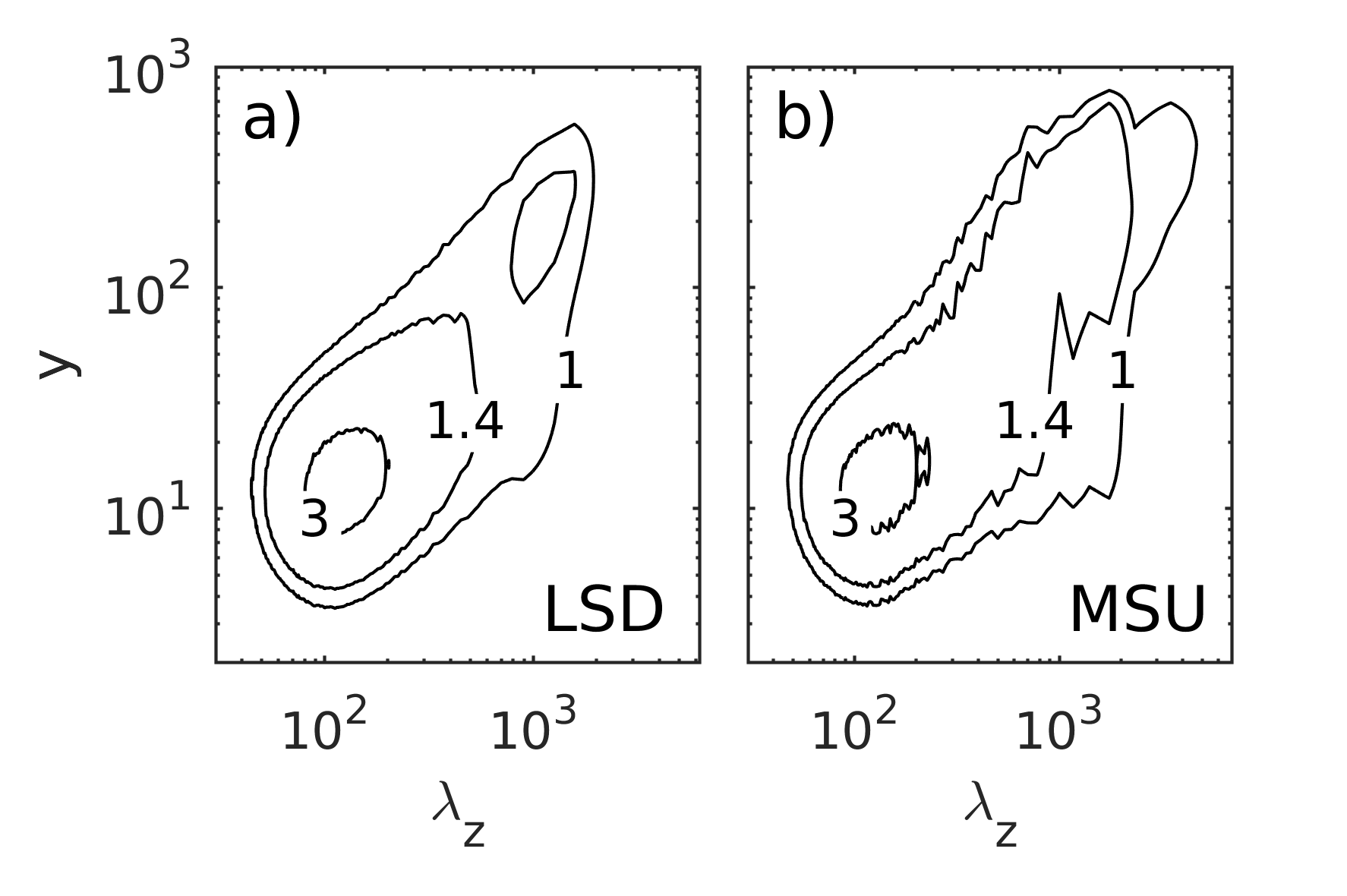}
  \caption{Premultiplied spanwise spectra $k_z h \, \Phi_{uu}$ of streamwise velocity fluctuations for a channel flow at $Re_\tau=1000$. Panel (a) refers to the LSD, while (b) to the MSU case. Viscous units. \label{fig:spectra}}
\end{figure}

\section{Results}

\begin{figure}[t]
  \centering
  \subfigure[$C_{AM}^\ast$-map, LSD. Viscous units.\label{fig:cam_lsd}]{%
    \includegraphics{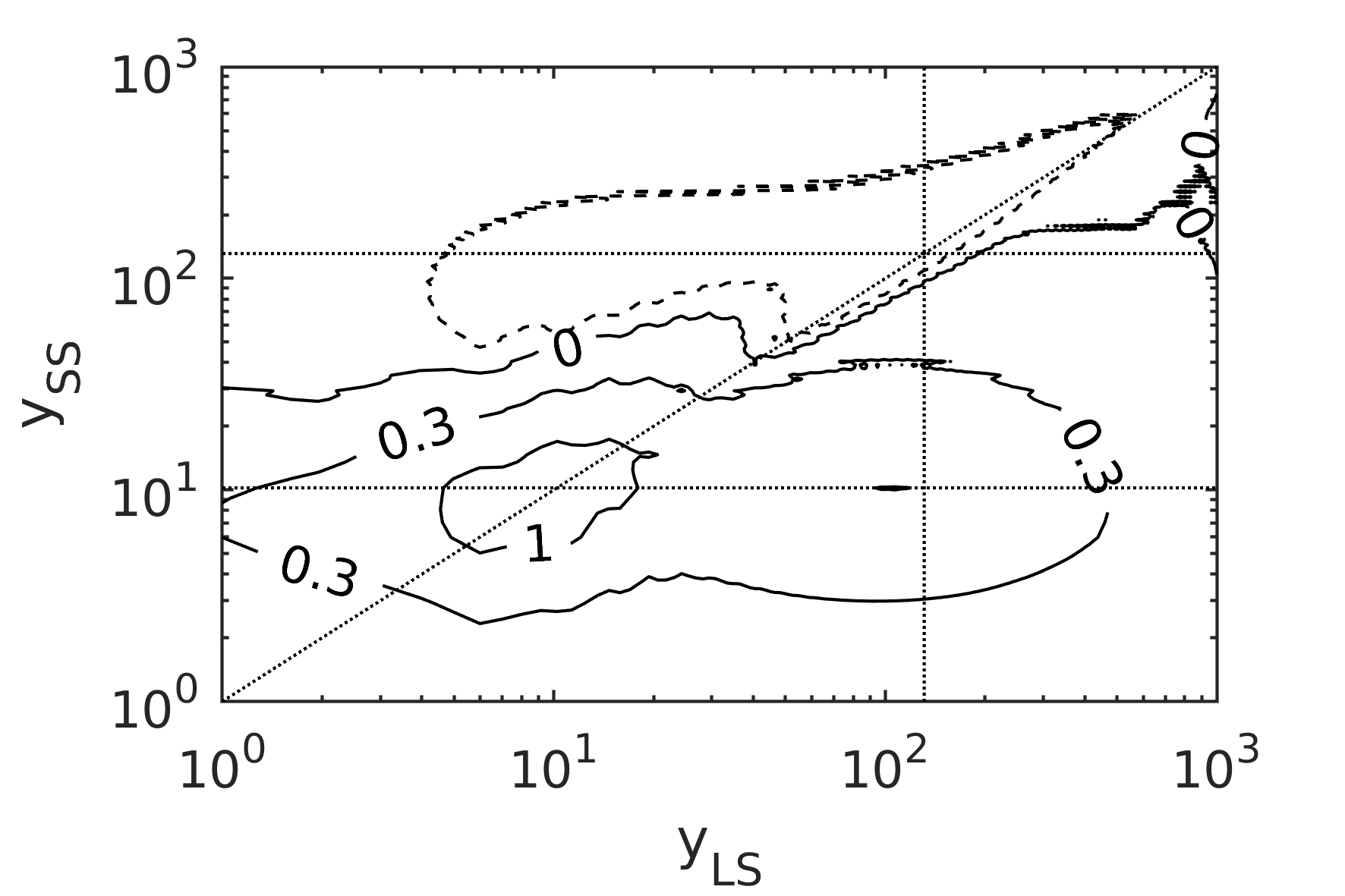}
  }
  \subfigure[$C_{AM}^\ast$-map, MSU. Viscous units.\label{fig:cam_msu}]{%
    \includegraphics{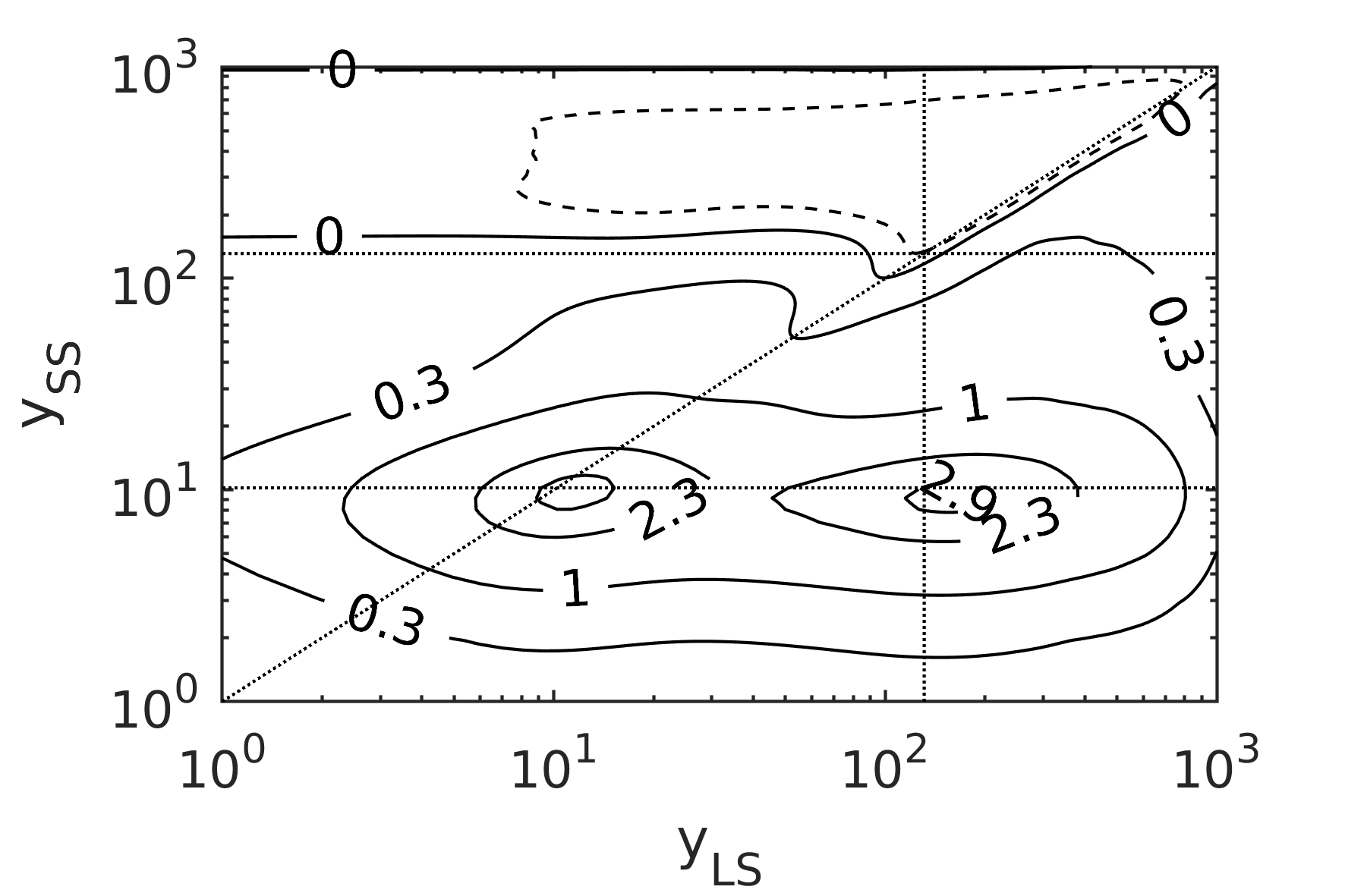}
  }
  \caption{$C_{AM}^\ast$-map for the channel flows at $Re_\tau=1000$. Panel \ref{fig:cam_lsd} refers to LSD, while \ref{fig:cam_msu} to MSU.}
  \label{fig:cam}
\end{figure}

The MSU simulation is validated in figures \ref{fig:spectra} and \ref{fig:cam} against the corresponding LSD case, with same $Re_\tau$ and $L_z$, but different streamwise domain length $L_x = 4 \pi h$. Panels \ref{fig:spectra}(a-b) show the premultiplied spanwise spectrum $k_z h \,\Phi_{uu}$ of streamwise velocity fluctuations; as expected, the weak, secondary peak of the spectrum at $y>100$ and $\lambda_z \approx 1000$ of the LSD case is significantly amplified in the MSU domain. Panels \ref{fig:cam_lsd} and \ref{fig:cam_msu} contain maps of the amplitude-modulation indicator $C_{AM}^\ast$. Qualitative agreement for the positive amplitude modualtion region is found for the MSU and LSD cases, suggesting that the MSU is also representative of the physics of a larger channel flow; once again, peaks of $C_{AM}^\ast$ are more pronounced in the MSU case as a consequence of the increased large-scale activity. Both cases are in agreement with the results of both \cite{bernardini-pirozzoli-2011} and \cite{agostini-etal-2016}, although the latter uses a correlation instead of the skewness here chosen as $C_{AM}^\ast$. Differently from the latter, we report the presence of two separate peaks as in \cite{bernardini-pirozzoli-2011} and \cite{eitel-amor-etal-2014} - albeit this difference might indeed be caused by the choice of indicator $C_{AM}^\ast$. The first diagonal peak lies at $y_{LS} \approx y_{SS} \approx 10$, whereas the second one lies at $y_{SS} \approx 10$, $y_{LS} \approx 100$. The main difference between the MSU and LSD cases is given by the boundaries of the outer-layer negative amplitude modulation region. This has an almost straight contour at $y_{SS} \approx 100$ in the MSU case; interestingly, such boundary happens to be at the same wall-normal distance as the large-scale position of the off-diagonal peak. By contrast, the same contour appears to be more rugged in the LSD case, possibly as a consequence of the meandering of the large scales - which is absent in a MSU.

\begin{figure}[t!]
  \centering
  \subfigure[$C_{AM}^+$-map (positive events, or sweeps). Viscous units.\label{fig:cam_plus}]{%
    \includegraphics{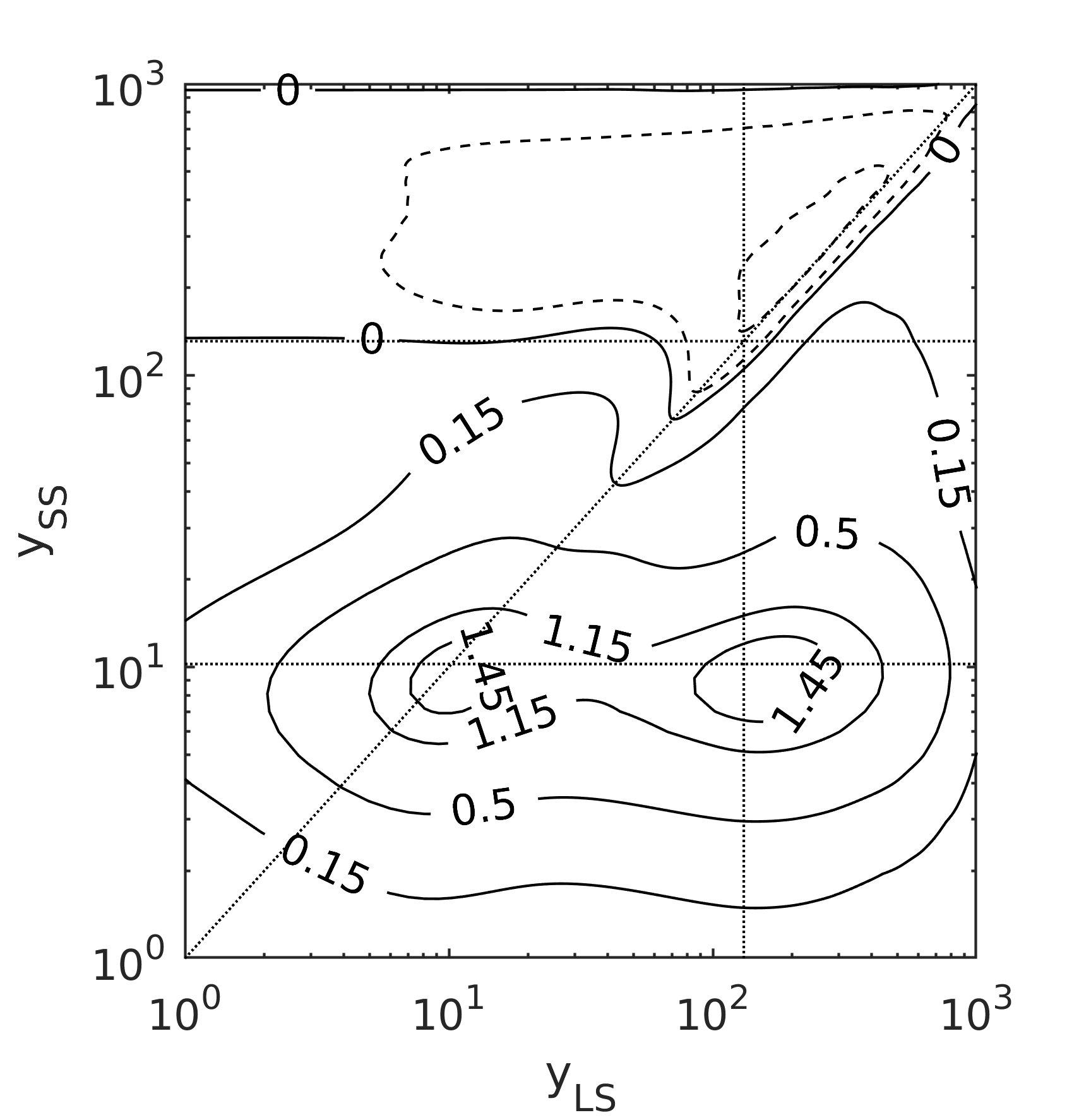}
  }
  \subfigure[$C_{AM}^-$-map (negative events, or ejections). Viscous units.\label{fig:cam_minus}]{%
    \includegraphics{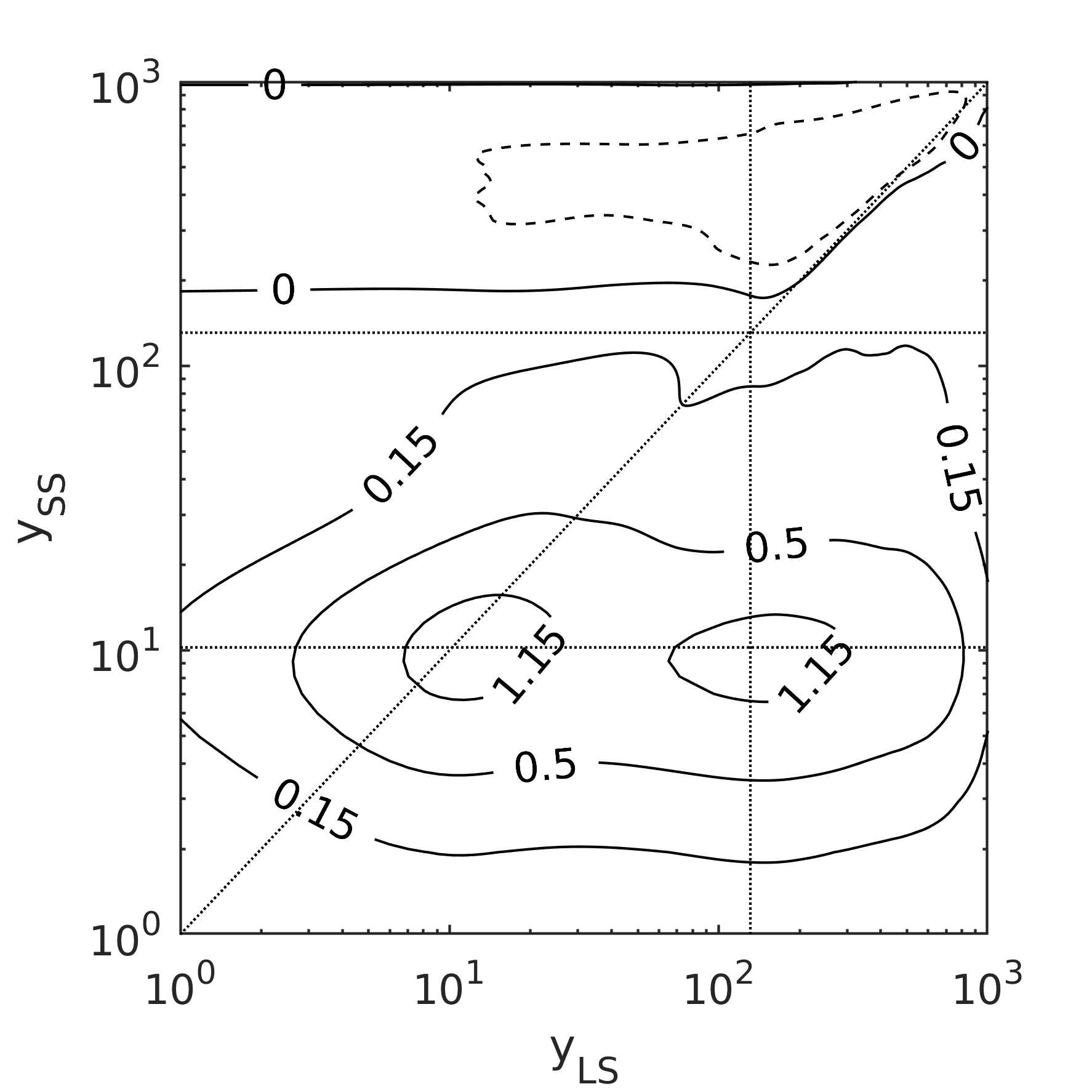}
  }
  \caption{Conditional maps of the amplitude modulation indicator $C_{AM}^\ast$ for a MSU channel flow at $Re_\tau=1000$.}
  \label{fig:cond_cam}
\end{figure}

Figure \ref{fig:cond_cam} reports the conditional amplitude-modulation covariances for the MSU as of equation \eqref{eq:camplus} and \eqref{eq:camminus}. Both positive and negative large-scale events (sweeps and ejections) contribute to both the inner, positive modulation and the outer, negative correlation regions. This is in agreement with the conjecture of \cite{agostini-leschziner-2019}, stating that modulation is an indirect phenomenon mediated by the gradient of the large scales. Indeed, they found that an increased large-scale gradient locally increases the production of small scales, and vice-versa, thus yielding the modulating effect. Also, the large-scale signal of a channel flow tends to have the same sign as its gradient in the proximity of the wall, and opposite sign at the channel center, this being valid for both sweeps and ejections: the reversal in sign of modulation in the outer layer is hence explained. This local interpretation surely is reasonable for the diagonal peak of $C_{AM}^\ast$, although its validity has to be verified for the non-diagonal one due to the non-local nature of the latter. 

\begin{figure}[t!]
  \centering
  \subfigure[$\left\langle u_{SS}^2(y_{SS}) u_{SS}^2(y_{LS}) \,|\, u_{LS}(y_{LS}) > 0 \right\rangle$. Viscous units.\label{fig:us4_plus}]{%
    \includegraphics{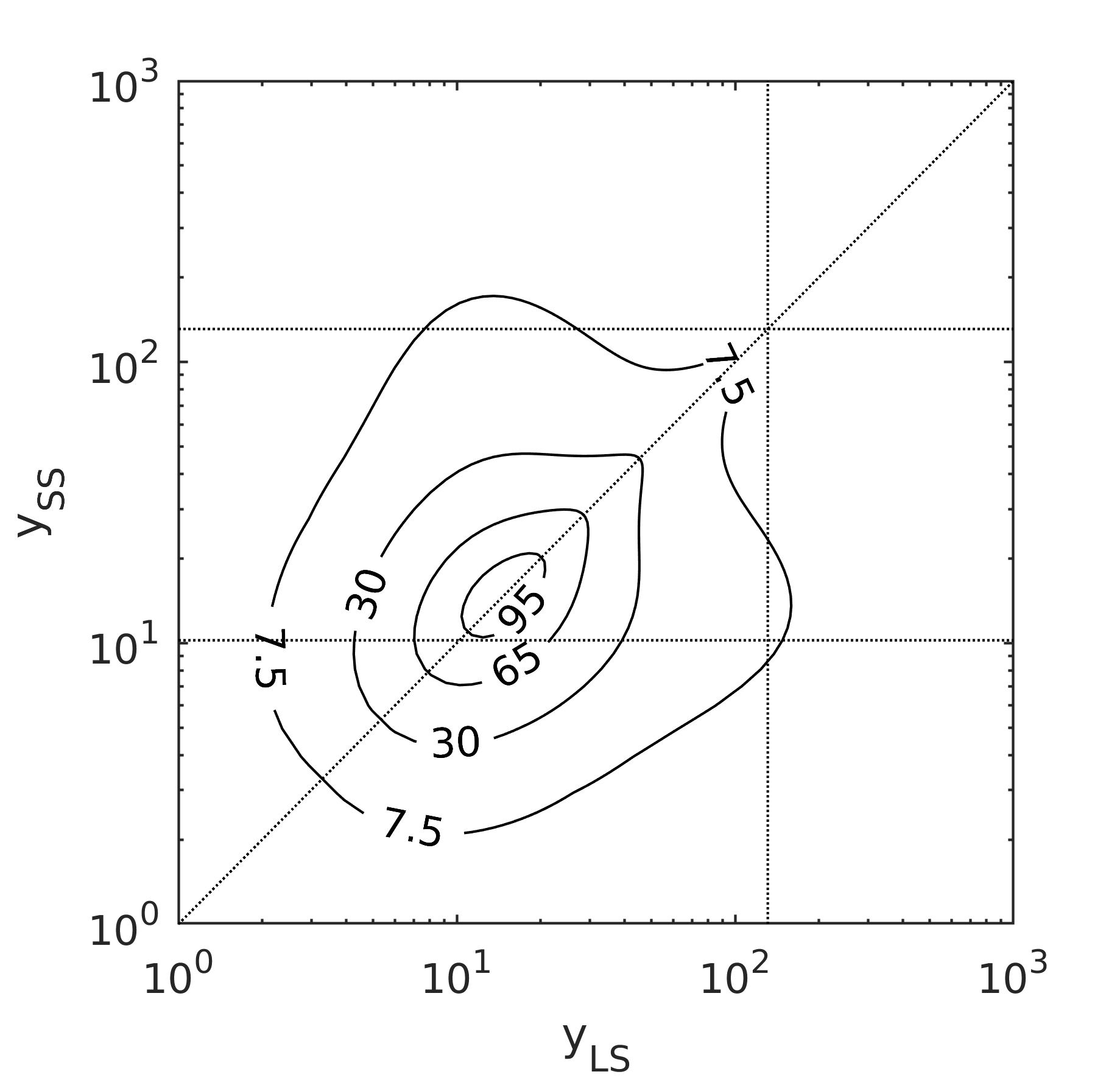}
  }
  \subfigure[$\left\langle u_{SS}^2(y_{SS}) u_{SS}^2(y_{LS}) \,|\, u_{LS}(y_{LS}) < 0 \right\rangle$. Viscous units.\label{fig:us4_minus}]{%
    \includegraphics{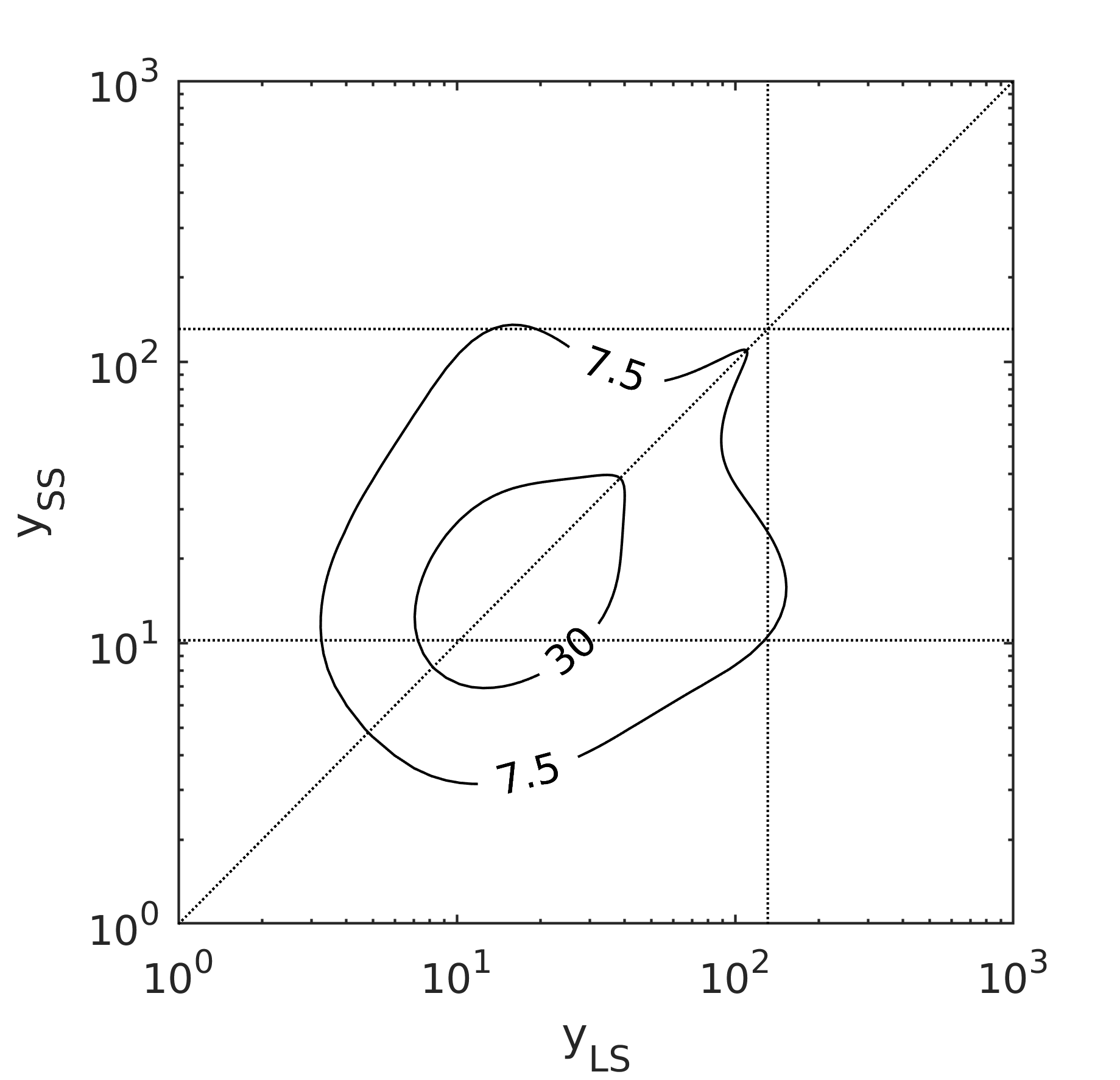}
  }
  \caption{Two-points conditioned autocovariance maps of $u_{SS}^2$ for a MSU channel flow at $Re_\tau=1000$.}
  \label{fig:cond_us4}
\end{figure}

The current data corroborates the idea of amplitude modulation as an asymmetric phenomenon \citep{agostini-leschziner-2014}, as sweeping events (figure \ref{fig:cam_plus}) seem to provide a stronger contribution to modulation in both the inner and the outer regions, whereas ejection events (figure \ref{fig:cam_minus}) have more confined maxima and minima. However weak, the positive-modulation region associated with ejection events is larger: for instance, the straight, horizontal portion of the zero-modulation isoline shifts from $y_{SS} \approx 130 $ in figure \ref{fig:cam_plus} to a farther  wall-normal position $y \approx 180$ in figure \ref{fig:cam_minus}. This effect is even more evident on the diagonal, where the negative-correlation region penetrates up until $y_{SS} \approx y_{LS} \approx 70$ for sweeping events, whereas it remains above $y_{SS} \approx y_{LS} \approx 200$ for ejections. 

A different way of detecting amplitude modulation is presented in figure \ref{fig:cond_us4}, which shows the two-point autocovariance of the $u_{ss}^2$ signal conditioned on the sign of the large-scales at point $y_{LS}$. Surprisingly, both the results in figures \ref{fig:us4_plus} and \ref{fig:us4_minus} appear to be symmetric with respect to the diagonal line, in spite of the fact that conditioning is done with respect to the $y_{LS}$ value only. Sweeping events are associated with an increased small-scale activity as can be seen on the diagonal of figure \ref{fig:us4_plus}; conversely, ejections are associated with reduced small scale activity (figure \ref{fig:us4_minus}). This is once again the exact idea behind amplitude modulation. Not only is the small-scale signal more intense in case of sweeps, but the peak in its intensity also spreads over a wider segment of the diagonal. Higher off-diagonal values can also be noticed in case of sweeps, meaning that the small scale envelopes at two different wall-normal positions correlate better with respect to the case of ejections. Hence the coherence of the envelope of small scales increases as sweeps occur, while it decreases in case of ejections.

\begin{figure}[t]
  \centering
  \subfigure[$\left\langle u_{LS}^2(y_{SS}) u_{LS}^2(y_{LS}) \,|\, u_{LS}(y_{LS}) > 0 \right\rangle$. Viscous units.\label{fig:ul_plus}]{%
    \includegraphics{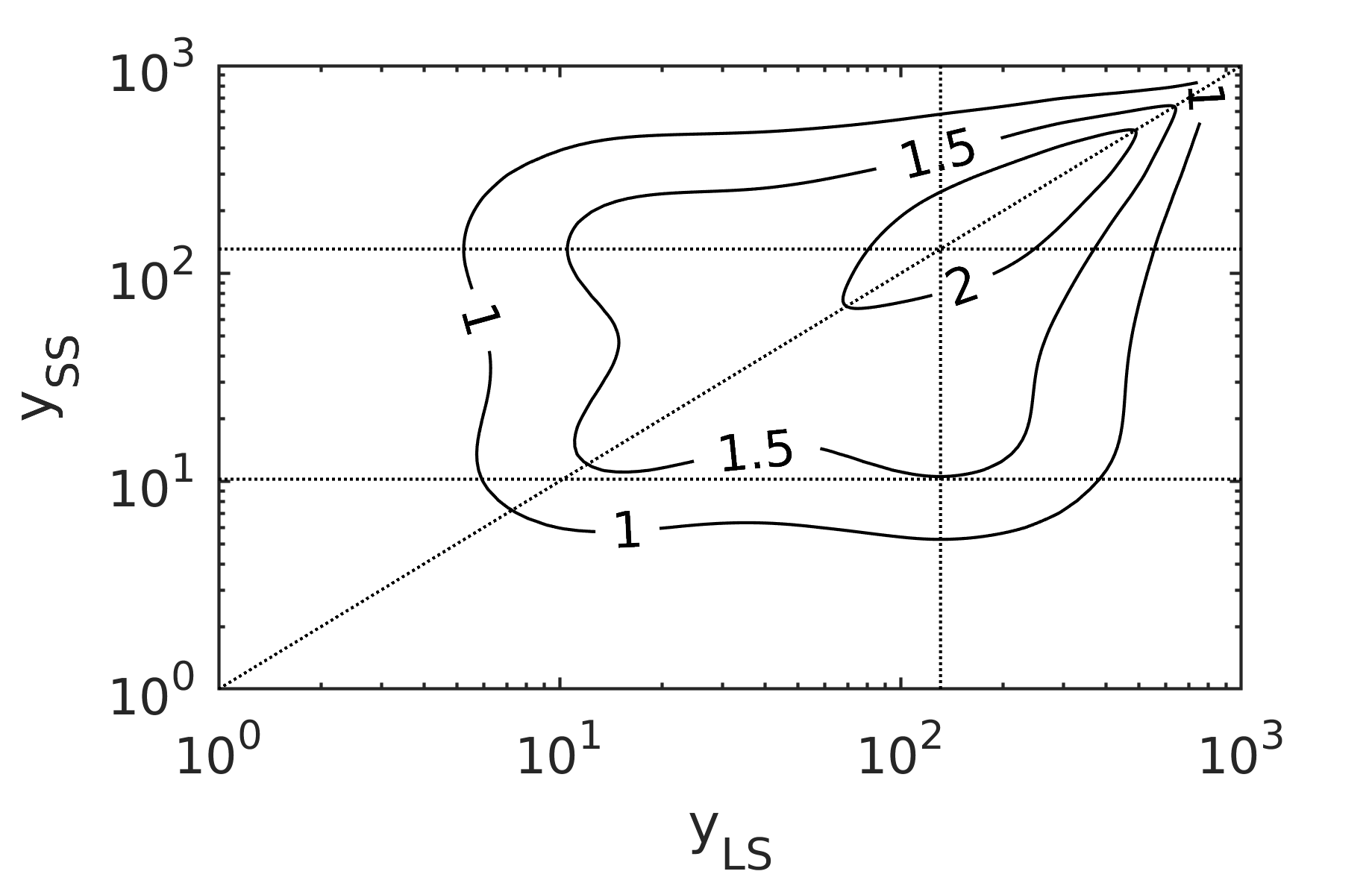}
  }
  \subfigure[$\left\langle u_{LS}^2(y_{SS}) u_{LS}^2(y_{LS}) \,|\, u_{LS}(y_{LS}) < 0 \right\rangle$. Viscous units.\label{fig:ul_minus}]{%
    \includegraphics{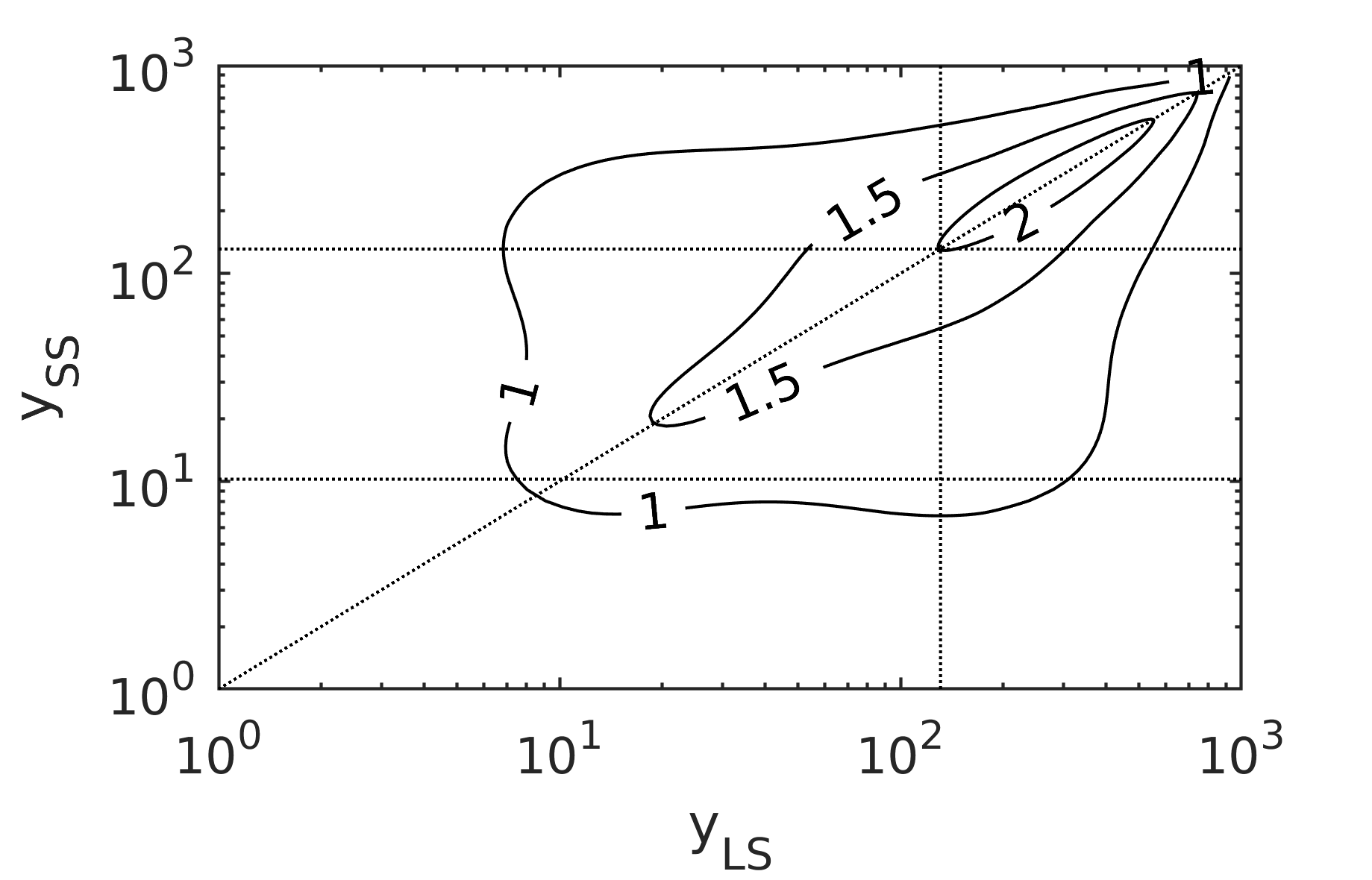}
  }
  \caption{Two-points conditioned autocovariance maps of $u_{LS}$ for a MSU channel flow at $Re_\tau=1000$.}
  \label{fig:cond_ul}
\end{figure}

As can be seen on the diagonal of both panels in figure \ref{fig:cond_us4}, small-scale activity peaks (as expected) in the buffer layer, roughly at the same position $y_{SS} \approx 10$ of the diagonal $C_{AM}^\ast$ peak. Conversely, large-scale activity has a much more elongated peak, which has its maximum in the outer layer and penetrates down to the buffer layer. This can be observed in figure \ref{fig:cond_ul}, showing the two-point autocovariance of the large scale fluctuation $u_{LS}$, once again conditioned on the sign of the large scales at $y_{LS}$. Hence the off-diagonal $C_{AM}^\ast$ peak corresponds to a $y_{SS}$ value of peaking small-scale activity, and to a $y_{LS}$ of high intensity of the large-scales: the presence of the $C_{AM}^\ast$ peak might thus be a consequence of peaking small- and large-scale activities. Nevertheless, strong, positive modulation is observed in such off-diagonal region also when normalising $C_{AM}^\ast$ with the magnitude of $u_{SS}^2$ and $u_{LS}$ \citep{agostini-etal-2016}, meaning that the peak is present as a consequence of matching phases between the signals (or, in other words, amplitude modulation).

As a final comment, large scales show an increased wall-normal coherence in correspondence of sweeping events, as highlighted by the high off-diagonal values of figure \ref{fig:ul_plus}; moreover, higher large-scale intensities are reached with respect to sweeping events (figure \ref{fig:ul_minus}). Fluctuations associated with sweeps also seem to penetrate deeper in the near-wall region, albeit marginally.

\section{Conclusions}

Amplitude modulation was here investigated in a turbulent channel flow in a minimal streamwise unit; such reduced computation domain is found to be representative of the large-small scale interactions of a larger, regular domain. Minor differences in the $C_{AM}^*$-maps of the two domains are possibly due to the meandering of large-scales, which is absent in the small domain.

By conditionally averaging the two-point amplitude-modulation covariance $C_{AM}^\ast$, it is found that both sweeps and ejections contribute to the two inner, positive-modulation peaks and to the outer, negative-modulation region. Sweeps provide more intense contributions to both positive- and negative-modulation regions, whereas the $C_{AM}^\ast$-map corresponding to ejections is flatter with narrower peaks. The inner, positive-modulation region is seen to reach farther towards the outer layer when the flow is subject to ejections; this is more evident on the diagonal of the amplitude-modulation map, where the negative-modulation region shrinks significantly. The differences between sweeping and ejection events, associated with opposite signs of the large-scale signal, corroborate the idea that amplitude modulation is an asymmetric phenomenon.

Finally, the off-diagonal peak of $C_{AM}^\ast$ is found to correspond to locations of high intensity for both large- and small-scale signals. An alternative way of investigating amplitude modulation consists in conditionally computing the autocovariance of the $u_{SS}^2$ signal; positive large-scale events correspond to increased small-scale activity, and vice versa.
\Acknowledgments

The authors acknowledge support by the state of Baden-W\"urttemberg through bwHPC as well as the Lundeqvist foundation. This work is supported by the Priority Programme SPP 1881 Turbulent Superstructures of the Deutsche Forschungsgemeinschaft.

\References
\vspace{-10pt}
\renewcommand{\bibsection}{} %
\setlength{\bibsep}{1pt} %
\bibliographystyle{etmm_bibstyle}
\bibliography{main.bib}
\end{document}